\documentclass[aps,prl,numerical,linenumbers,superscriptaddress,showpacs,floatfix,reprint]{revtex4-1}
\usepackage[dvipdfm]{graphicx}
\usepackage{amsmath,amssymb,amsfonts}
\usepackage{color}
\usepackage[colorlinks=true,linkcolor=blue,plainpages=false,hypertex]{hyperref}

\def\v2{\mbox{$v_2$}}

\newcommand{\mean}[1]{\left\langle #1 \right\rangle}

\begin{document}
%

\title{ Azimuthal anisotropy: transition from hydrodynamic flow to jet suppression
}
%
%
\author{ Roy~A.~Lacey}
\email[E-mail: ]{Roy.Lacey@Stonybrook.edu}
\affiliation{Department of Chemistry, 
Stony Brook University, \\
Stony Brook, NY, 11794-3400, USA}
\affiliation{Physics Department, Bookhaven National Laboratory, \\
Upton, New York 11973-5000, USA}
\author{A. Taranenko}
\affiliation{Department of Chemistry, 
Stony Brook University, \\
Stony Brook, NY, 11794-3400, USA}
\author{ R. Wei}
\affiliation{Department of Chemistry, 
Stony Brook University, \\
Stony Brook, NY, 11794-3400, USA}
\author{ N. N. Ajitanand} 
\affiliation{Department of Chemistry, 
Stony Brook University, \\
Stony Brook, NY, 11794-3400, USA}
\author{ J. M. Alexander}
\affiliation{Department of Chemistry, 
Stony Brook University, \\
Stony Brook, NY, 11794-3400, USA}
\author{ J. Jia}
\affiliation{Department of Chemistry, 
Stony Brook University, \\
Stony Brook, NY, 11794-3400, USA}
\affiliation{Physics Department, Bookhaven National Laboratory, \\
Upton, New York 11973-5000, USA}
%
%
%
\author{\\ R. Pak}
\affiliation{Physics Department, Bookhaven National Laboratory, \\
Upton, New York 11973-5000, USA}
\author{Dirk H. Rischke}
\affiliation{Frankfurt Institute for Advanced Studies (FIAS), Frankfurt am Main, Germany} 
\affiliation{Institut f\"ur Theoretische Physik, Johann Wolfgang Goethe-Universit\"at \\
             D–-60438 Frankfurt am Main, Germany} 

\author{ D. Teaney} 
\affiliation{Department of Physics, 
Stony Brook University, \\
Stony Brook, NY, 11794-3800, USA}
\affiliation{Physics Department, Bookhaven National Laboratory, \\
Upton, New York 11973-5000, USA}

\author{ K. Dusling} 
\affiliation{Physics Department, Bookhaven National Laboratory, \\
Upton, New York 11973-5000, USA}

\date{\today}


\begin{abstract}
 
	Measured $2^{\text{nd}}$ and $4^{\text{th}}$ azimuthal anisotropy 
coefficients $v_{2,4}(N_{\text{part}}, p_T)$ are scaled with 
the initial eccentricity $\varepsilon_{2,4}(N_{\text{part}})$ of the 
collision zone and studied as a function of the number of participants 
$N_{\text{part}}$ and the transverse momenta $p_T$. 
Scaling violations are observed for $p_T \alt 3$ GeV/c, consistent with 
a $p_T^2$ dependence of viscous corrections and a linear increase of 
the relaxation time with $p_T$. These empirical viscous corrections to flow and 
the thermal distribution function at freeze-out constrain estimates of the specific 
viscosity 
and the freeze-out temperature 
for two different models for the initial collision geometry. The apparent 
viscous corrections exhibit a sharp maximum for $p_T \agt 3$ GeV/c, 
suggesting a breakdown of the hydrodynamic ansatz and the onset of a 
change from flow-driven to suppression-driven anisotropy.
\end{abstract}
\maketitle

	
	A central objective of the experimental program at the Relativistic
Heavy Ion Collider (RHIC) is to delineate the thermodynamic and transport 
properties of the hot and dense matter produced in energetic heavy ion 
collisions. This matter can equilibrate to 
form a hot plasma of quarks and gluons (QGP)\cite{Shuryak:1978ij} which 
rapidly expands, cools, and then hadronize
to produce the observed particles. 
The hydrodynamic-like expansion of the QGP, as well as its interactions 
with hard scattered partons, results in the anisotropic emission of hadrons 
relative to the reaction plane.  
At mid-rapidity, the magnitude of this momentum anisotropy is characterized by the 
even-order Fourier coefficients;
%
%
$
v_{\rm n} = \mean{e^{in(\Delta\phi)}}, {\text{  }} n=2,4,..., 
$
%
%
where $\Delta\phi$ is the azimuth of an emitted hadron 
about the reaction plane, and brackets denote 
averaging over particles and events.

	The coefficients for hadrons with low transverse momenta ($p_T \alt 2$ GeV/c)
can be understood in terms of {\it flow} or partonic interactions which drive pressure 
gradients in an initial ``almond-shaped" collision zone produced in non-central 
collisions \cite{Ollitrault:1992bk,Heinz:2001xi,Shuryak:2008eq,Peschanski:2009tg}. 	 
	For higher transverse momenta ($p_T \agt 5$ GeV/c) the coefficients 
can be attributed to jet quenching \cite{Gyulassy:1993hr} -- the 
process by which hard scattered partons interact and lose energy in the hot and 
dense QGP, prior to fragmenting into hadrons. This energy 
loss manifests as a suppression of hadron yields \cite{Adcox:2001jp} which
depends on the average distance that partons propagate through the QGP. 
Thus, $v_{2,4}$ stem from the fact 
that partons which traverse the QGP medium in a direction parallel (perpendicular) to the 
reaction plane result in less (more) suppression due to shorter (longer) 
parton propagation lengths \cite{Gyulassy:2000gk,Wang:2000fq,Liao:2008dk}. 
This path-length dependence is exemplified in the recently observed scaling 
patterns for hadron suppression \cite{Lacey:2009ps,Lacey:2009kg}. 
	The transition from flow-driven to suppression-driven anisotropy 
is still poorly understood, and it remains a challenge to find a single 
consistent theoretical framework that gives
an explanation of $v_{2,4}$ measurements over the full $p_T$ range. 

    The magnitude of $v_{2,4}$, as well as their detailed dependence 
on $p_T$ and collision centrality (or number of participants $N_{\text{part}}$), 
give invaluable insights on the thermodynamic and transport coefficients 
of the QGP. 
In particular, flow measurements ($v_2(p_T)$ and $v_2(N_{\text{part}})$) 
have been used to estimate the specific shear viscosity ({\em i.e.},\ the ratio 
of shear viscosity $\eta$ to entropy density $s$ of the plasma) via comparisons 
to viscous relativistic hydrodynamic calculations \cite{Romatschke:2007mq,
Luzum:2008cw,Song:2008hj,Chaudhuri:2009hj,Denicol:2010tr}.  
%
	The reliability of these $\frac{\eta}{s}$ estimates 
is influenced not only by the uncertainties in the initial conditions for hydrodynamic evolution, 
but also by ambiguities in the departure from equilibrium on the freeze-out surface. 
For a viscous fluid, this distribution ($f$) is of the form
\begin{equation}
\frac{dN}{dy p_T dp_T d\phi} \sim f_0 + \delta f \equiv f_0\left( 1+ C\left(\frac{p_{T}}{T_{\!f}}\right)^{2-\alpha}\right),
%
\label{eq2}
\end{equation}  
where $f_0$ is the equilibrium distribution, $T_{\!f}$  is the freeze-out temperature, 
$C \approx \frac{\eta}{3\tau s T_{\!f}}$ and  $\alpha$ 
ranges between 0 and 1  \cite{Teaney:2003kp,Dusling:2009df};
$\tau$ is the time scale of the expansion.
The factor $\delta f$, which results from a finite shear viscosity, is known to 
dominate the calculated viscous corrections to $v_2(p_T)$ for 
$p_T \agt 1$ GeV/c \cite{Dusling:2009df}.
However, its momentum dependence and associated relaxation time $\tau_R(p)$ is 
not known {\it a priori}, and it is unclear whether 
or not it is proportional to $p_T^2$ ($\alpha =0$ and $\tau_R \propto p$) as has been 
generally assumed in hydrodynamic calculations \cite{Romatschke:2007mq,Luzum:2008cw,
Song:2008hj,Chaudhuri:2009hj,Dusling:2009df,Denicol:2010tr}. 
The freeze-out temperature and the $p_T$ for which large viscous corrections render a breakdown 
of viscous hydrodynamics are also not well established experimentally.

\begin{figure}[t]
\includegraphics[width=1.0\linewidth]{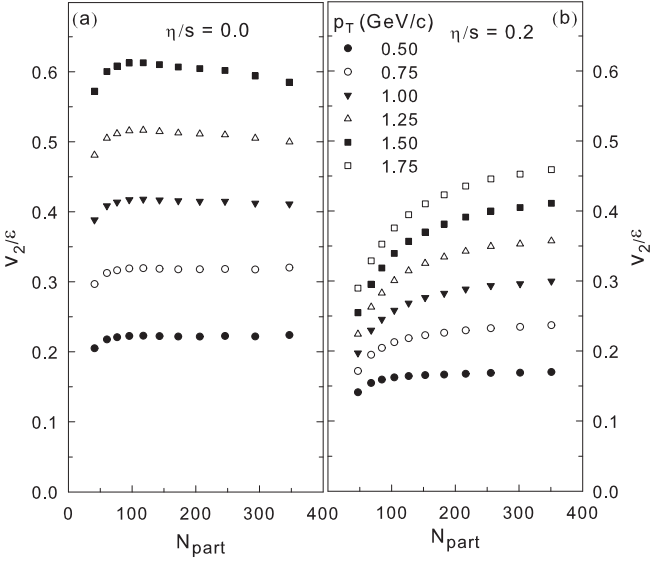} 
\caption{(color online) Comparison of $v_2/\varepsilon_2$ vs. $N_{\rm part}$ 
for several $p_T$ selections, obtained from perfect fluid (a) and 
viscous hydrodynamic (b) simulations \cite{Dusling:2009df} of Au+Au collisions.  
}
\label{Fig1}
\end{figure}

	The influence of viscous corrections on the eccentricity-scaled 
anisotropy coefficient $\frac{v_{2}(N_{\text{part}}, p_T)}{\varepsilon_{2}(N_{\text{part}})}$
is illustrated in Figs. \ref{Fig1} (a) and (b) where the results of hydrodynamic 
simulations (with the code of Dusling and Teaney \cite{Dusling:2009df}) are shown 
for $\frac{\eta}{s}=0$ and 0.2, respectively.
Fig. \ref{Fig1} (b) shows that viscous effects reduce $v_{2}(N_{\text{part}}, p_T)$ and break 
the scale invariance of ideal hydrodynamics evidenced in Fig. \ref{Fig1} (a), {\em i.e.} there 
are deviations away from the essentially flat $N_{\text{part}}$ dependence expected 
for ideal hydrodynamic scaling.
These deviations from eccentricity-scaling can be used to estimate and characterize
viscous corrections \cite{Bhalerao:2005mm,Drescher:2007cd,Song:2008si,Lacey:2009xx,Dusling:2009df}. 

	Here, we use the eccentricity scaled 
anisotropy coefficients,  
$\frac{v_{2}(N_{\text{part}}, p_T)}{\varepsilon_{2}(N_{\text{part}})}$ and 
$\frac{v_{4}(N_{\text{part}}, p_T)}{\varepsilon_{4}(N_{\text{part}})}$, 
to extract estimates of the viscous corrections 
to $v_{2,4}(p_T,N_{\text{part}})$. 
In turn, we use these estimates to explore the $p_T$ dependence of $\delta f$ and the 
transition from flow-driven to suppression-driven anisotropy.   
We find viscous correction factors for $p_T \alt 3$ GeV/c that 
validate the commonly assumed $p_T^2$ dependence of $\delta f$,   
and give constrained estimates for $\frac{\eta}{s}$ and 
the freeze-out temperature $T_{\!f}$.
For $p_T \agt 3$ GeV/c, the apparent viscous corrections 
signal a breakdown of the hydrodynamic ansatz.
 
%

	The $v_{2,4}$ data employed in our analysis are selected 
from the high-precision PHENIX measurements recently reported 
for Au+Au collisions at $\sqrt{s_{NN}}$ = 200 GeV \cite{Adare:2010ux}. 
These data show that both $v_{2}$ and $v_{4}$ have a strong dependence on  
$p_T$ and centrality. 
The large increase in $v_{2,4}(p_T)$ from central to peripheral 
events is especially important to our study of viscous corrections.
%
\begin{figure}[t]
\includegraphics[width=1.0\linewidth]{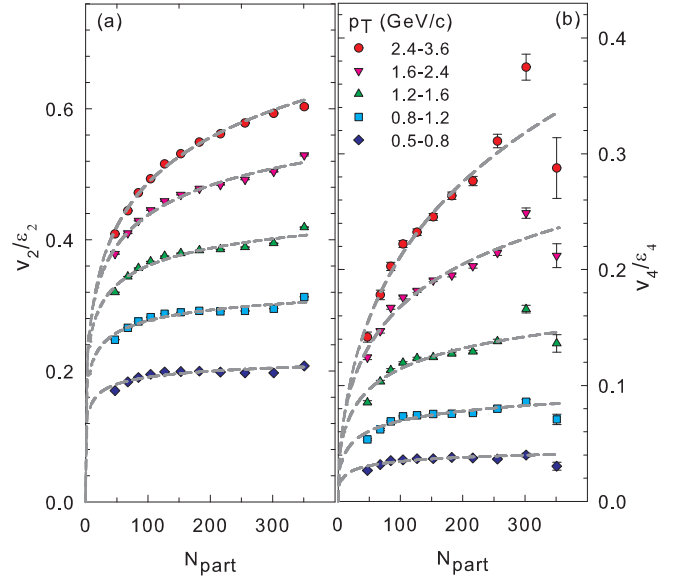} 
\caption{(color online) Comparison of $v_2/\varepsilon_2$ vs. $N_{\rm part}$ (a) 
and $v_4/\varepsilon_4$ vs. $N_{\rm part}$ (b) for several $p_T$ selections as 
indicated \cite{Adare:2010ux}. The dashed curves indicate a simultaneous fit to 
the data in (a) and (b) [for each $p_T$] with Eq. (\ref{eq3}). The $v_{2,4}$ data 
are from Ref. \cite{Adare:2010ux}. 
}
\label{Fig2}
\end{figure}

	Monte Carlo (MC) simulations \cite{Lacey:2010yg} were used to calculate the 
event-averaged geometric quantities used in our analysis. 
For each collision, the values for $N_{\rm part}$ and the number of binary 
collisions $N_{\text{coll}}$ were determined via the Glauber ansatz \cite{Miller:2007ri}. 
The associated values for the transverse size $\bar{R}$, area $S$ 
and eccentricities $\varepsilon_{2,4}$ were then evaluated from the 
two-dimensional density of sources in the transverse  
plane $\rho_s(\mathbf{r_{\perp}})$ using two principal models; a modified version of 
the MC-Glauber approach \cite{Miller:2007ri} and the factorized Kharzeev-Levin-Nardi (MC-KLN) 
model \cite{Lappi:2006xc,Drescher:2007ax}. 

	For each event, we compute an event-shape vector $S_{n}$ and the azimuth of 
rotation $\Psi_n^*$ for the $n$-th harmonic of the shape profile \cite{Lacey:2010yg}; 
  \begin{eqnarray}  
    S_{nx} & \equiv & S_n \cos{(n\Psi^*_n)} = 
    \int d\mathbf{r_{\perp}} \rho_s(\mathbf{r_{\perp}}) \omega(\mathbf{r_{\perp}}) \cos(n\phi), \label{eq:S_x} \nonumber \\
    S_{ny} & \equiv & S_n \sin{(n\Psi^*_n)} = 
    \int d\mathbf{r_{\perp}} \rho_s(\mathbf{r_{\perp}}) \omega(\mathbf{r_{\perp}}) \sin(n\phi), \label{eq:S_y} \nonumber \\
    \Psi^*_n & = & \frac{1}{n} \tan^{-1}\left(\frac{S_{ny}}{S_{nx}}\right), \label{eq:S_n-plane} \nonumber
  \end{eqnarray}
where $\phi$ is the azimuthal angle of each source and the 
weight $\omega(\mathbf{r_{\perp}}) = \mathbf{r_{\perp}}^2$;
$\varepsilon_{2,4}$ were calculated as:
\begin{eqnarray}
\varepsilon_2 = \left\langle \cos 2(\phi - \Psi^*_2) \right\rangle, \,\,\,
\varepsilon_4 = \left\langle \cos 4(\phi - \Psi^*_2) \right\rangle,
\label{e2e4}
\end{eqnarray}
where the brackets denote averaging over sources, as well as events belonging to 
a particular centrality or impact parameter range. For the MC-Glauber 
calculations, an additional entropy-density weight was applied reflecting the 
combination of spatial coordinates of participating nucleons and 
binary collisions  \cite{Hirano:2009ah};
%
\begin{eqnarray}
\rho_s(\mathbf{r_{\perp}}) \propto \left[ \frac{(1-\alpha_1)}{2}\frac{dN_{\text{part}}}{d^2\mathbf{r_{\perp}}} + 
                     \alpha_1 \frac{dN_{\text{coll}}}{d^2\mathbf{r_{\perp}}} \right], \nonumber
\label{Eq5}
\end{eqnarray}
where $\alpha_1 = 0.14$ was constrained by multiplicity measurements as a 
function of $N_{\text{part}}$ for Au+Au collisions \cite{Back:2004dy}.
Note that $\varepsilon_{2,4}$ (cf. Eq. (\ref{e2e4})) correspond 
to $v_{2,4}$ measurements in the so-called participant 
plane \cite{Alver:2006wh}; this is analogous to the measurement of 
$v_{2,4}$ with respect to the $2^{\text{nd}}$ order event-plane as
described in Ref. \cite{Adare:2010ux}. 
A correlation between the principal axes of the quadrupole ($\Psi^*_2$) and 
hexadecapole ($\Psi^*_4$) density profiles 
was also introduced to 
account for contributions to $v_4$ from $v_2$ \cite{Lacey:2010yg}. 
This correlation has a    
significant influence only on $\varepsilon_4$ in the most 
central collisions \cite{Lacey:2010yg}.


%
\begin{figure}[t]
\includegraphics[width=0.75\linewidth]{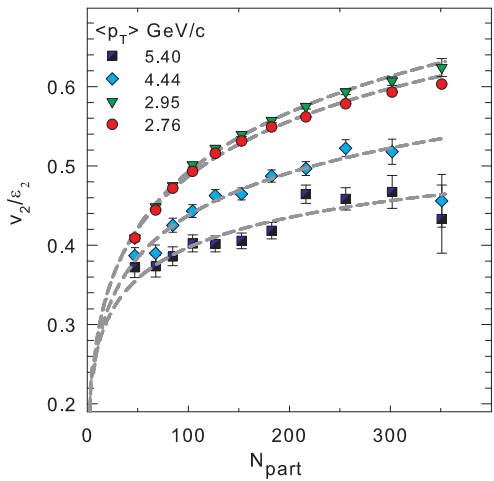}
\caption{(color online) $v_2/\varepsilon_2$ vs. $N_{\rm part}$  
for several $\left< p_T \right>$ values as indicated.  
The filled circles are the same as in Fig. \ref{Fig2}(a).
The dashed curves show fits to the data obtained with Eq. (\ref{eq3}).
The $v_{2}$ data are from Ref. \cite{Adare:2010ux}. 
}
\label{Fig3}
\end{figure}

		Figures \ref{Fig2} and \ref{Fig3} show eccentricity-scaled values of 
$v_{2,4}(p_T,N_{\text{part}})$ obtained with MC-KLN eccentricities for several $p_T$ 
cuts. The low-$p_T$ selections are almost flat, {\it i.e.} small scaling violations.
These violations gradually increase with $p_T$ over 
the $p_T$ range indicated in Fig. \ref{Fig2}. That is,  
the data points slope upward progressively (from low to high $N_{\rm part}$) 
as the $\left< p_T \right>$ is increased. Fig. \ref{Fig3} shows that
this trend reverses to give scaling violations which decrease with 
increasing $\left< p_T \right>$, for $\left< p_T \right> \agt 3$ GeV/c.
This inversion could be an indication for the onset of suppression-driven 
anisotropy as discussed below.
Similar scaling performed with MC-Glauber eccentricities,  
show the same trends exhibited in Figs. \ref{Fig2} and \ref{Fig3}, albeit 
with larger scaling violations, as discussed below. 

    In lieu of detailed model comparisons \cite{comp2010}, we estimate the magnitude 
of the viscous corrections by parametrizing the observed scaling violations 
with a Knudsen number ($K ={\lambda}/{\bar R}$) 
ansatz akin to that in Refs. \cite{Bhalerao:2005mm,Drescher:2007cd};
%
\begin{equation}
 \frac{v_{2k}(p_T)}{\varepsilon_{2k}} = \frac{v_{2k}^{\rm{h}}(p_T)}{\varepsilon_{2k}} 
  \left[ \frac{1}{{1+ \frac{K^*(p_T)}{K_0}}} \right]^{k}\; k=1,2,...\;,	
	\label{eq3}
\end{equation}
where $K^*(p_T)$ characterizes 
the magnitude of the viscous correction for a given $p_T$, $\frac{v_{2,4}^{\rm{h}}(p_T)}{\varepsilon_{2,4}}$ 
are the eccentricity-scaled coefficients expected from ideal hydrodynamics,   
$\lambda$ is the mean-free path, 
and $K_0$ is a constant estimated to be $0.7 \pm 0.03$ with the aid of a transport 
model \cite{Gombeaud:2007ub}. 

	For each $p_T$ selection, $[K^*(p_T)]^{-1} = \beta(p_T) \frac{1}{S}\frac{dN}{dy}$; 
($\frac{dN}{dy} \propto N_{part}$)
is evaluated by fitting  
$\frac{v_{2,4}(p_T,N_{\text{part}})}{\varepsilon_{2,4}(N_{\text{part}})}$ vs. $N_{\text{part}}$ 
with Eq. (\ref{eq3}) (cf. curves in Figs \ref{Fig2} and \ref{Fig3}).  
The fit parameters $\beta(p_T)$, so obtained, allow the determination of $K^*(p_T)$ as a 
function of $N_{\rm part}$.
Note that a model uncertainty in the value of $K_0$ would lead to an accompanying 
uncertainty in the magnitude of $K^*$. However, the consistency of our procedure 
with hydrodynamic models has been tested via fits to 
$\frac{v_{2}(p_T,N_{\text{part}})}{\varepsilon_{2}(N_{\text{part}})}$ vs. $N_{\text{part}}$, 
obtained for specified values of $\frac{\eta}{s}$ \cite{Lacey:2009xx}.
These fits lead to $\frac{\eta}{s}$ values which 
reproduce the input values to the hydrodynamic simulations.

%
%
\begin{figure}[t]
\includegraphics[width=1.0\linewidth]{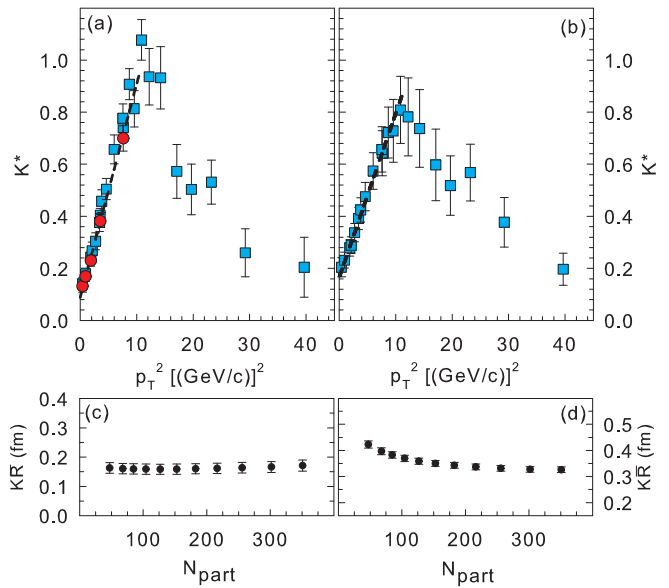} 
\caption{(color online)  $K^*$ vs. $\left< p_T \right>^2$ (a) and (b),  
and $K\bar{R}$ vs. $N_{\rm part}$ (c) and (d), extracted with 
MC-KLN (left panels) and MC-Glauber geometry (right panels).
The filled circles in (a) indicate results from the simultaneous fits shown 
in Fig. \ref{Fig2}. The dashed curves in (a) and (b) show a fit to the data
for $\left< p_T \right>^2 \alt 10$ [GeV/c]$^2$.
}
\label{Fig4}
\end{figure}

	Figures \ref{Fig4} (a) and (b) show the values of 
$K^*$ vs. $p_T^2$, extracted for $N_{\rm part} \sim 351$ with MC-KLN 
and MC-Glauber geometries respectively (the plots for other values of 
$N_{\rm part}$ show similar trends, but with different intercepts).
The filled circles in Fig. \ref{Fig4} (a) show results from the simultaneous fits indicated 
by the dashed curves in Figs. \ref{Fig2} (a) and (b). The squares show results for fits 
which employ only $\frac{v_2(p_T,N_{\text{part}})}{\varepsilon_2(N_{\text{part}})}$ 
data  \cite{Adare:2010ux}; both are in good agreement. 

	The $K^*$ values shown in Figs. \ref{Fig4} (a) and (b) indicate 
a linear dependence on $p_T^2$, for $\left< p_T \right>^2 \alt 10$ [GeV/c]$^2$, 
that demonstrates a non-zero viscosity and the $p_T^2$ dependence of $\delta f$ commonly 
assumed in hydrodynamic simulations \cite{Romatschke:2007mq,Luzum:2008cw,Song:2008hj,Chaudhuri:2009hj,
Dusling:2009df,Denicol:2010tr}. In contrast, the data for 
$10 \alt \left< p_T \right>^2 \alt 40$ [GeV/c]$^2$ show a striking trend inversion. 
We interpret this as a signal for the breakdown of the hydrodynamic ansatz when 
$K^* \sim 1$, as well as an indication for the onset of suppression-driven anisotropy. 
Note that such a scenario would lead to improved eccentricity scaling for 
$p_T \agt 3$ GeV/c because the eccentricity encodes the variation of 
the path length relative to the orientation of the reaction plane, and the 
suppression of hadron yields has been found to increase as $\frac{1}{\sqrt{p_T}}$
for a similar $p_T$ range \cite{Lacey:2009kg}.

	To obtain estimates for $K$ and $T_{\!f}$ for each value of $N_{\text{part}}$, 
we use our observation that $K^*(p_T) \propto p_T^2$ in conjunction with 
the first order expansion of 
$
v_2(p_{T}) =  \langle \cos(2\Delta\phi) \rangle_{p_T}
    \equiv 
    \frac{
     \int^{\pi}_{-\pi} d\Delta\phi \,\cos(2\Delta\phi)  \,
                \frac{d^3N }{ dy\, p_t\,dp_t\,d\Delta\phi}
        }{
       \int^{\pi}_{-\pi}\, d\Delta\phi\,  
                \frac{d^3N }{ dy\,p_t\,dp_t\,d\Delta\phi}  
        }\;,
$
to obtain the expression $K^*(p_T) = K + \frac{B}{T_{\!f}} \left(\frac{p_T}{T_{\!f}}\right)^2$;
the constant $B$ was cross-checked via fits to the results from hydrodynamic simulations.
Fits to $K^*$ vs. $p_T^2$ were performed   
with this fit function. 
%
The dashed curve in Fig. \ref{Fig4}(a) indicates such a fit 
for $\left< p_T \right>^2 \alt 10$ [GeV/c]$^2$; it gives the values 
$K=0.09 \pm 0.01$ (from the intercept) and $T_{\!f} = 162 \pm 11$ MeV
(from the slope). 
The same fit to the $K^*$ values in Fig. \ref{Fig4}(b) 
(extracted with MC-Glauber eccentricities), give the values $K=0.17 \pm 0.007$ 
and $T_{\!f} = 173 \pm 11$ MeV. These same values of $T_{\!f}$ are indicated by   
the fits to the data for other $N_{\text{part}}$ values, for both data sets. 
However, as to be expected, the extracted $K$ values vary with $N_{\text{part}}$.
Note again that a model uncertainty in the value of $K_0$ would lead to an accompanying 
uncertainty in the magnitude of $K$ and the associated values for $\lambda$ and 
$\frac{\eta}{s}$ discussed below.
The estimates for $T_{\!f}$ are similar to the chemical freeze-out temperature
($T_c \sim 165$ MeV) obtained for a broad range of collision energies \cite{Cleymans:2005xv}.

	Figures \ref{Fig4} (c) and (d) show the product 
$K\bar{R}$ vs. $N_{\text{part}}$, obtained with MC-KLN and MC-Glauber geometry 
respectively. Fig. \ref{Fig4} (c) indicates that the estimated value 
$\lambda \sim 0.17 \pm 0.018$ fm is essentially independent of $N_{\text{part}}$.
Fig. \ref{Fig4} (d) indicates a larger estimate for central collisions 
$\lambda \sim 0.33 \pm 0.02$ fm, and a mild increase as collisions 
become more peripheral. While our analysis seems more consistent 
for the MC-KLN geometry, the model dependencies apparent in Fig. \ref{Fig4}, highlight 
the importance of experimental signatures that can distinguish MC-KLN and MC-Glauber 
collision geometries \cite{Lacey:2010yg}.

	Estimates for $\frac{\eta}{s}$ were obtained via the expression 
$\frac{\eta}{s} \approx \lambda T c_s \equiv (\bar{R}KT c_s)$ where the 
sound speed $c_s = 0.47 \pm 0.03$~c was obtained from lattice 
calculations \cite{Huovinen:2009yb} for the mean 
temperature $T=220 \pm 20$ MeV \cite{Adare:2008fqa}. This gives
the estimates $4\pi\frac{\eta}{s} = 1.1 \pm 0.1$ 
and $4\pi\frac{\eta}{s} = 2.1 \pm 0.2$ for the K values extracted 
using MC-KLN and MC-Glauber eccentricities [respectively] in central and 
mid-central collisions. 
These estimates are in agreement with the low value from prior 
extractions~\cite{Lacey:2006bc,Adare:2006nq,Romatschke:2007mq,Luzum:2008cw,
Xu:2007jv,Drescher:2007cd,Song:2008hj,Chaudhuri:2009hj,
Lacey:2009xx,Dusling:2009df,Denicol:2010tr}. 
%


In summary, we have used eccentricity scaled anisotropy coefficients 
to extract estimates of the strength and role of the viscous corrections. 
These estimates show a quadratic increase with  
$p_T$ (for $p_T \alt 3$ GeV/c) that validates a non-zero viscosity and 
a relaxation time which grows with $p_T$. 
The extracted viscous corrections also constrain the estimates 
$4\pi\frac{\eta}{s} \sim 1.1 \pm 0.1 \;(2.1 \pm 0.2)$ and 
$T_{\!f} = 162 \pm 11 \;\text{MeV} \; (173 \pm 11 \;\text{MeV})$ 
for MC-KLN (MC-Glauber) collision geometries for a strongly coupled plasma. 
The onset of a transition from 
flow-driven to suppression-driven anisotropy is signaled by a  
sharp maximum of the apparent viscous corrections 
for $p_T \agt 3$ GeV/c. These results provide valuable constraints for 
input parameters to more detailed viscous hydrodynamic calculations.

{\bf Acknowledgments}
This research is supported by the US DOE under contract DE-FG02-87ER40331.A008 and 
by the NSF under award number PHY-0701487.
 


%
\bibliography{delta_f_refs} 

\begin{thebibliography}{39}
\bibitem [{com()}]{comp2010}%
  \BibitemOpen
  \href@noop {} {}\bibinfo {note} {Such comparisons are in progress and will be
  presented elsewhere.}\BibitemShut {Stop}%
\end{thebibliography}
\end{document}